\documentclass{iopart}

\usepackage{amssymb}
\usepackage{amsthm}

 \usepackage{epsfig}

\newcommand{\sect}[1]{\setcounter{equation}{0}\section{#1}}

\def\be{\begin{equation}}
\def\ee{\end{equation}}
\def\bea{\begin{eqnarray}}
\def\eea{\end{eqnarray}}

\def\p{\partial}
\def\a{\alpha}

\def\g{\gamma}
\def\G{\Gamma}
\def\e{\varepsilon}
\def\k{\kappa}
\def\l{\lambda}
\def\m{\mu}
\def\o{\omega}

\def\z{\zeta}

\def\vfi{\varphi}

\def\ra{\rangle}
\def\la{\langle}

\def\CS{{coherent }}

\newcommand{\Sc}{Schr\"odinger }

\begin{document}

 \large

\title {Darboux transformations of coherent states of the
 time-dependent singular oscillator}

\author{
 Boris F Samsonov
}
\address{Physics Department of Tomsk State
 University, 634050 Tomsk, Russia}

\address{ Departamento de F\'{\i}sica Te\'orica, Universidad de
Valladolid,  47005 Valladolid, Spain}

\ead{\mailto{samsonov@phys.tsu.ru}}

\begin{abstract}
\baselineskip=16pt
\noindent
Darboux transformation of both Barut-Girardello and Perelomov
coherent states for the time-dependent singular oscillator is
studied. In both cases the measure that realizes the resolution of
the identity operator in terms of coherent states is found and
corresponding holomorphic representation is constructed. For the
particular case of a free particle moving with a fixed value of the
angular momentum equal to two it is shown that Barut-Giriardello
coherent states are more localized at the initial time moment
while the Perelomov coherent states are more stable with respect
to time evolution. It is also illustrated that
Darboux transformation may keep unchanged this different time behavior.
\end{abstract}

 \submitto{\JPA}

\medskip
\medskip



\medskip
\medskip

\textbf{Corresponding Author}:

B F Samsonov

Physics Department

Tomsk State University

36 Lenin Ave

634050 Tomsk

RUSSIA

\medskip

Phone: 7 3822 913 019

E-mail: {\it samsonov@phys.tsu.ru\/}


\newpage

\sect{Introduction}

There exist several definitions of \CS  states
(CS)
\cite{CS}-\cite{GK}.
They lead to the same result for the harmonic oscillator
potential and usually to different results for other physical
systems. Nevertheless, a careful analysis shows that they
possesses common properties which can be taken
to define CS for the very general quantum system \cite{GK}.
It happens that such a definition is very ambiguous.
Therefore one can choose
 between different possible systems the one which
has a desirable property. For instance,
one can look for a state, the wave function of which can
 be expressed in terms of known special functions
or even in terms of elementary functions only.
This simplifies considerably the study of these
CS and their use in other applications. In this way one was
able to construct different systems of CS for the
well-known soliton potentials \cite{SSP}. It was also shown that
 different CS may be related between them with a symmetry operator
 \cite{SSP} or with the Laplace transform  \cite{BVM}.

The most essential progress in studying CS has been achieved
for systems possessing symmetries \cite{Per}. If it is not
possible to associate a symmetry group with a quantum system the
problem becomes much more complicated. For instance,
after applying a nontrivial
supersymmetric (or equivalently Darboux)
transformation to a Hamiltonian allowing a symmetry,
 the symmetry is usually lost or it is transformed to a nonlinear
 symmetry \cite{SJMP}.
 This makes impossible to apply group
 theoretical methods for studying CS. In
 this respect it was suggested to apply the same transformation to
 the known CS of the initial quantum system and treat states
 thus obtained as coherent states for the transformed system
 \cite{DCS,BSDCS}.
 This approach proved to be useful for studying
CS for supersymmetrically transformed Harmonic oscillator
 \cite{BSDCS} and time-independent singular oscillator \cite{BSSO}
 potentials. In particular, using correspondence between
 classical and quantum systems, which can be realized just by
 the technique of CS (see e.g. \cite{Per}),
 one was able to give an interpretation of the
 supersymmetry transformation in terms of classical notions
 \cite{SJMP}. Here we continue this study at the level of the
 time-dependent singular oscillator.

The time-dependent
singular oscillator Hamiltonian
\be\label{h0}
h_0=-\p_x^2+\o^2(t)x^2+gx^{-2}
\ee
plays an essential role in different physical applications
between which we would like to mention interesting
results in
molecular physics \cite{mp}, in optics \cite{opt} and
in mathematical physics \cite{mph}.
Recently a model of a
two-ion trap has been proposed
(see \cite{DMR} where a good literature review is also given)
based on this Hamiltonian.
Exact solvability of the \Sc equation with the
Hamiltonian (\ref{h0})
 may be, in particular, related with the
fact that the symmetry operators  realize a representation
of the $su(1,1)$ algebra and therefore it can be solved by the
method of separation of variables \cite{M}.
Moreover,
for this algebra two
systems of CS are commonly known \cite{DMR}.
The one may be obtained with
the help of the $SU(1,1)$ group translation operator
(they are known as Perelomov CS, see e.g. \cite{Per})
and the other are
eigenstates of the annihilation operator \cite{AC}
(Barut-Girardelo CS \cite{BG}). Here we will study the
supersymmetric transformation of either system.

The paper is organized as follows.
In the next section to fix the notations
we give a review of some properties of the singular oscillator
Hamiltonian and its CS.
In particular, basing on the Barut-Girardello CS
 we construct a holomorphic representation of state functions
 and operators; for a concrete example of a free particle
moving with a fixed value of the angular momentum
equal to two
 we compare the time stability of Barut-Girardello and Perelomov
 CS.
In section 3 we apply the time-dependent Darboux transformation
both to Barut-Girardello and to Perelomov CS and study some
properties of states thus obtained.
Conclusions are drown in the last section.

\section{Singular oscillator Hamiltonian}

In this section we summarize briefly what is known about
this Hamiltonian and its \CS states 
(see e.g. \cite{SJMP},\cite{BSSO}-\cite{DMR},\cite{AC},\cite{BG})
 we need further.
We would like to stress that
although almost everything presented here is already
published, but
from one side
we are adopting here a different approach
with respect to the one usually used
and from the other side
we are bringing together a number of facts spread in numerous
literature.
In contrast to most of papers
 using different modifications of the method of quantum
invariants \cite{inv} for solving the \Sc equation
we are using the method of separation of
variables \cite{M}.
The advantage of this approach is that we do not need to
be restricted by any quantum mechanical picture.
Instead, we consider
the \Sc equation as a second order parabolic differential equation, which
gives us the possibility to get ``nonphysical" solutions we need
for applying the Darboux algorithm.

\subsection{Solutions of the \Sc equation}

Generators of the $SU(1,1)$ symmetry group, being symmetry
operators
of the \Sc equation with the Hamiltonian $h_0$,
 in coordinate representation
\bea\label{kpmk}
k_{-}=2\left[ a^2-\e ^2gx^{-2}\right] \qquad
k_{+}=2\left[ \left( a^{+}\right) ^2-
\overline{\e }^2gx^{-2}\right],
\\ \label{k0k}
k_0=\textstyle{\frac 12}\left( k_{-}k_{+}-k_{+}k_{-}\right)
\eea
are expressed in terms of the harmonic oscillator creation and
annihilation operators
\[
a=\e \partial _x-\textstyle{\frac i2}\dot \e x\qquad
a^{+}=-\bar{\varepsilon }\partial _x+
\textstyle{\frac i2}\dot {\bar{\e }}x\,.
\]
The dot over a symbol means the derivative with respect to time.
Parameters $\e$ and $\bar\e$ are solutions
to the equation of motion for
the classical harmonic oscillator
\[
\ddot \varepsilon +4\omega ^2(t)\varepsilon =0\,.
\]
In particular, when $\e$ is complex, $\bar\e$ will denote its
complex conjugate and they are such that
$\dot\e\bar\e-\e\dot{\bar\e}=\frac 12i$.
Casimir operator $C$ is expressed in terms of the parameter $g$:
$C=\frac 12\left( k_{+}k_{-}+k_{-}k_{+}\right) -k_0^2=3/16-g/4$.
This gives us the relation between $g$ and the representation
parameter $k$: $C=k(1-k)$, $g=3/4+4k(k-1)$.

Solutions $\psi_n(x,t)$, $n=0,1,\ldots$ of the \Sc equation
 belonging to the Hilbert space
of functions square integrable on semiaxis form
 a basis of the discrete series
irreducible representation $T_k^+(g)$ of the group
 $SU(1,1)$ at $k=\frac 12+\frac 14\sqrt{1+4b}$. The vacuum vector
  $\psi_0(x,t)$ is selected from the set of solutions
  of the \Sc equation by  the equations
 $k_0\psi_0(x,t) =k\psi_0(x,t)$, $k_{-}\psi_0(x,t) =0$.
All other basis vectors may be obtained by acting with the
creation operator $k_+$:
\[
\psi_n(x,t) =
(-1)^n\sqrt{\frac{\Gamma \left( 2k\right) }{n!\Gamma \left(
n+2k\right) }}\left( k_{+}\right) ^n\psi_0(x,t) \,.
\]
The action of the operators (\ref{kpmk})-(\ref{k0k})
on this basis is given by
\bea\label{kmin}
k_0\psi_n(x,t)=\left( k+n\right) \psi_n(x,t)\qquad
k_\pm\psi_n(x,t) =-c_n^\pm\psi_{n\pm 1}(x,t)\\
c_n^\pm=(n+\textstyle{\frac 12}\pm \textstyle{\frac 12})^{\frac 12}
(n+2k-\textstyle{\frac 12}\pm \textstyle{\frac 12})^{\frac 12}\,.
\nonumber
\eea

To solve the \Sc equation simultaneously with the eigenvalue
equation for $k_0$ we are using the method of separation of
variables. From coordinates $(x,t)$ we go to curvilinear
coordinates $(u,v)$ defined as: $u=x\g^{-1/2}$, $v=t$ were $\g=\e
\bar\e$.
The choice
\be\label{psiPQ}
\psi=e^{\frac i8u^2\dot\g}P(v)Q(u)
\ee
guarantees the separation of variables.
The eigenvalue equation for $k_0$ is reduced to the first order
ordinary equation for $P$
which can readily be integrated to give
$P=\g^{-1/4}\left( {\bar\e}/{\e} \right)^\l $.
So, replacing the
action of $k_0$ on a solution by the multiplication on the
separation constant $\l$ one obtains from the \Sc equation the
following second order differential equation:
\be
Q_{yy}-\left[ \textstyle{\frac 14}y^2+ gy^{-2}-2\l \right]Q=0\,\quad
y=\textstyle{\frac 12}u\,.
\ee
This equation can be reduced to the equation for the Laguerre
polynomials if $\l=n+k$, $n=0,1,\ldots$, which just corresponds to
the well-known coordinate representation of the basis
$\psi_n(x,t)$.
If $\l=-k-n$ or $\l=k-n-1$, $n=0,1\ldots$ it can give rise to
Laguerre polynomials also
but of course this does not correspond to square
integrable solutions of the \Sc equation.
Finally these solutions have the form
\be\label{psin}
\psi_n=N_{0n}\g^{-1/4}\left( \frac{\bar\e}{\e} \right)^\l
y^{\a +\frac 12} e^{(\pm\frac 14 +\frac i2 \dot\g)y^2}
L^{\a}_n\left(\mp\textstyle{\frac 12}{y^2}\right)\qquad
y=\textstyle{\frac 12}(\e\bar\e)^{-{\frac 12}}x\,.
\ee
The lower sign and the choice
$\a=2k-1$ correspond to the discrete spectrum eigenfunctions.
The upper sign permitting $\a$ to
be both positive and negative
corresponds to solutions which do not belong to the Hilbert space.
For $\a=2k-1$ they are such that $k_0\psi_n(x,t)=-(k+n)\psi_n(x,t)$
and for $\a=1-2k$ they satisfy the equation
 $k_0\psi_n(x,t)=(k-n-1)\psi_n(x,t)$.
 The value
$N_{0n}= 2^{3k-{\frac 12}}(n!)^{{\frac 12}}\G^{-{\frac 12}}(n+2k)$
guarantees the normalization of the square integrable solutions
 to the unity.
For the non-normalizable solutions $N_{0n}$ does not play any role.

For applying the Darboux algorithm we need nodeless solutions. The
zeros of the Laguerre polynomials are well-known \cite{Beitmen}.
 In the
negative semiaxis, $x<0$, $L_n^\a(x)$
has only one node provided
$(\a +1)_n=(\a+1)(\a+2)\ldots (\a +n)<0$.
It is clear that since we put
$k=\frac 12+\frac 14\sqrt{1+4b}$, $\a=2k-1=\frac 12\sqrt{1+4b}>0$
meaning that in this case any $\psi_n$ is nodeless and
it is such that
$\psi_n^{-1}(\infty)=0$ but $\psi_n^{-1}(0)=\infty$. For a negative
$\a$ the result depends on the interval where $\a$ falls.
If $-2m-1>\a >-2m-2$ we have $n=0,2,\ldots ,2m$ and for
$-2m\ge \a\ge -2m-1$ more values of $n$ are possible:
$n=0,2,\ldots ,2m, 2m+1, 2m+2,\ldots$ where $m=0,1,\ldots $.
Moreover, for $\a <-3/2$ the function
$\psi_n^{-1}(x)$ is square integrable and
belongs to the domain od definition of $h_0$ when it is considered
as an operator in the Hilbert space.

\subsection{Barut-Girardelo coherent states}

The states $\psi_\m(x,t)$ known as Barut-Girardelo CS are defined as eigenstates
of the operator $k_-$
\begin{equation}\label{Km}
k_-\psi_\m(x,t)=\m\psi_\m(x,t)\, \quad \m\in {\Bbb C}\,.
\end{equation}
To solve this equation simultaneously with the \Sc
equation we are using the method of separation of
variables once again.
In coordinates $\left\{u=x/\e, v=t\right\}$ the \Sc equation
separates if $\psi_\m=\exp(\frac 14 \e\dot\e u^2)P(v)Q(u)$. The
function $P(v)$ is defined by a first order ordinary equation
which is easily integrated to give
\[
P=\e^{-{\frac 12}}e^{-2\m^2\frac{\bar\e}{\e}}\,.
\]
The \Sc equation is reduced to the second order equation for $Q$:
\be\label{Q}
Q''(u)-(gu^{-2}+\m^2)Q(u)=0
\ee
which after the change of the dependent variable $Q=u^{{\frac 12}}I$
becomes the modified Bessel equation.
The condition for
$\psi_\m(x,t)$ to belong to the domain of definition of $h_0$
selects us only one solution to this equation: $I=I_{2k-1}(\m u)$.
(We are using the standard notations for the Bessel functions
\cite{Beitmen}.)
To calculate the normalization integral we are
making use of the tables
\cite{Prudnikov}.
So,  the normalized solution is
\be\label{psimu}
\psi_\m(x,t)=\textstyle{\frac{1}{2\e}}I_{2k-1}^{-{\frac 12}}(4\m\bar\m)
\sqrt x\,
I_{2k-1}\left( \textstyle{\frac{1}{\e }}\m x\right)
\exp({\frac{i\dot\e}{4\e} x^2-\frac{2\bar\e}{\e}\m^2})\,.
\ee
Expanding the Bessel function in a Taylor series one can get their
expansion in terms of Laguerre polynomials and in such a way
they are expressed
through the basis functions (\ref{psin}) with coefficients depending
only on even powers of $\m$. Therefore, it is more convenient to
change the complex variable $\m $ in favor of $\l=2\m^2$ which
yields
\be\label{fcs}
\psi_\l(x,t)=N_{0\l}\sum_{n=0}^\infty a_n\l^n\psi_n(x,t)\qquad
a_n=\frac{(-1)^n\sqrt{\G(2k)}}{\sqrt{n!\G(n+2k}}
\ee
with
\be
N_{0\lambda }:= \la \psi_0 | \psi_\l \ra =
\left(\bar{\lambda }\lambda \right) ^{\frac k2-\frac 14}
I_{2k-1}^{-{\frac 12}}\left( 2\left|
\lambda \right| \right) \Gamma ^{-{\frac 12}}\left( 2k\right)\,.
\ee
As usual the states $\psi_\l(x,t)$ are not orthogonal to each
other
\[
\langle \psi_{\lambda ^{^{\prime }}}\mid \psi_{\lambda} \rangle =
\frac{I_{2k-1}\left(
2\sqrt{\vphantom{I^{I^I}}\lambda \bar{\lambda }^{{}^\prime}}\right) }
{\left[ I_{2k-1}\left(
2\left| \lambda \right| \right) I_{2k-1}\left( 2\left| \lambda ^{^{\prime
}}\right|
\right) \right] ^{{\frac 12}}}\,.
\]

Since the basis $\psi_n(x,t)$ is complete in the Hilbert space one
can calculate the measure $\rho(\l)$ which realizes
 the resolution of the identity over the coherent states
 $\psi_\l$ \cite{BSSO}
\be\label{Ip}
\int |\psi_\l\ra\la \psi_\l |d\rho(\l) =1\qquad
\rho(\l)=\textstyle{\frac 1\pi} K_{2k-1}(2|\l |)I_{2k-1}(2|\l |)d\l d\bar\l
\,.
\ee
All integrals over the variable $\l$ are extended to the whole
complex plan.

Now one can construct a holomorphic representation of the
vectors and operators \cite{BSSO}.
Any $\psi(x,t)=\sum_{n=0}^{\infty}c_n\psi_n(x,t)$ from the
Hilbert space of square integrable functions defined on the
positive semiaxis
can be written in coherent state representation $\psi^c(\l)$:
\be
\psi ^c\left( \l \right) :=\langle \psi_{\bar{\lambda }}(x,t)\mid \psi(x,t) \rangle
={N}_{0{\lambda }}
\sum_{n=0}^\infty a_nc_n\lambda ^n\equiv
{N}_{0{\lambda }}\psi \left( \lambda \right) \quad  \lambda \in
\Bbb C\,.
\ee
The holomorphic function $\psi(\l)=\sum_{n=0}^\infty a_nc_n\lambda ^n$
can be associated with the usual square integrable function
$\psi(x,t)$ given by its Fourier coefficients $c_n$ over the basis
(\ref{psin}).

Using the complex conjugate form of the
 resolution of the identity (\ref{Ip}) one can define an inner
product
$\langle \psi _a\left( \lambda \right) \mid
\psi _b\left( \lambda \right) \rangle$
in the space of the functions $\psi_{a,b}(\l)$ holomorphic in the complex
plane
\be\label{sp}
\langle \psi _a\mid \psi _b\rangle
=\int \left| N_{0\lambda }\right|^2
\bar{\psi }_a\left( \lambda \right) \psi _b\left( \lambda
\right) d\rho \left( \lambda \right) := \langle \psi _a\left( \lambda \right) \mid
\psi _b\left( \lambda \right) \rangle \,.
\ee
This means that the integration in the space of holomorphic
functions should be carried out with the measure
$d\tilde\rho \left( \lambda \right)=|N_{0\l}|^2d\rho(\l)$,
so that
\be\label{hip1}
\langle \psi _a\left( \lambda \right) \mid \psi _b\left( \lambda \right)
\rangle =\int \overline{\psi }_a\left( \lambda \right) \psi _b\left( \lambda \right)
d\tilde\rho \left( \lambda \right) \,.
\ee
To distinguish this inner product from the one in the space
$L^2(0,\infty )$ we indicate  the integration variable inside the
brackets.
The space of holomorphic functions $\psi(\l)$ such that
\[
\int|\psi(\l)|^2d\tilde\rho(\l)<\infty
\]
equipped with the inner product (\ref{hip1}) becomes a Hilbert space.

The orthonormal basis $\psi_n(x,t)$ in this representation looks like as
follows:
 \[
 \psi _n\left( \lambda \right) = a_n\lambda ^n \qquad
 \langle \psi _n\left( \lambda \right)\mid \psi _{n^{\prime }}
\left( \lambda \right) \rangle =\delta _{nn^{\prime }}\,.
\]
From here we find the Dirac-delta function
\[
\delta \left( \lambda ,\lambda ^{\prime }\right) =\sum_{n=0}^\infty\psi _n\left( \lambda
\right) \overline{\psi }_n\left( \lambda ^{\prime }\right)
=\sum_{n=0}^\infty a_n^2
( \lambda \overline{\lambda }^{\prime }) ^n=\Gamma \left( 2k\right)
( \lambda \overline{\lambda }^{\prime }) ^{1-2k }I_{2k-1}
\left( 2\sqrt{\lambda \overline{\lambda }^{\prime }}\right)
\]
which takes off the integration
in this space:
\[
\psi (\l) =\int \delta \left( \lambda ,\lambda ^{\prime
}\right) \psi (\l ')d\tilde\rho \left( \lambda ^{\prime }\right).
\]
The CS in the holomorphic representation
\[
\psi_{\l '}(\l)=(\l\l ')^{\frac 12-k}I_{2k-1}(2\sqrt{\l\l '})
\]
and
the $SU(1,1)$ generators
\begin{equation}
k_0=\lambda \frac d{d\lambda }+k
\qquad k_{+}=\lambda
\qquad k_{-}=\lambda \frac{d^2}{d\lambda ^2}+2k\frac d{d\lambda } \,.
\label{KG}\end{equation}
 are time-independent.
The mean values of the operator $k_0$ in the coherent state
\begin{equation}\label{E0}
\langle \psi_\lambda \left| k_0\right| \psi_\lambda
\rangle =k+\left| \lambda \right| \frac{I_{2k+1}\left( 2\left| \lambda \right|
\right)
}{I_{2k-1}\left( 2\left| \lambda \right| \right) }
\end{equation}
and
\be
\langle \psi_\lambda \left| k_0^2\right| \psi_\lambda
\rangle =k^2+\left| \lambda \right| ^2+\left| \lambda \right|
\frac{I_{2k+1}\left(2\left| \lambda \right| \right) }
{I_{2k-1}\left( 2\left| \lambda \right| \right) }\
\ee
may be useful in the following.

\subsection{Perelomov coherent states}

Perelomov CS $\psi_z(x,t)$ are obtained by acting on the ground state
function with the group translation operator. Therefore their
Fourier coefficients over the basis $\psi_n(x,t)$ are independent
on $t$ and coincide with the ones for the time-independent Hamiltonian
\bea\label{csf}
\psi_z(x,t)=N_{0z}\sum_{n=0}^\infty a_n\,z^n\,\psi_n(x,t)\qquad |z|<1\\
N_{0z}=(1-|z|^2)^k\qquad
a_n=\sqrt{\frac{\G(n+2k)}{n!\G(2k)}}~.
\eea
Using the generating function for the Laguerre polynomials one
gets their explicit expression
\be\label{psiz}
\psi_z(x,t)=2^{\frac 12-3k}\G^{-\frac 12}(2k)
\e^{-2k}x^{2k-\frac 12}
\left(\frac{1-|\z|^2}{(1-\z)^{2}}\right)^k
\exp\left(
-\frac{x^2}{16\g}\frac{1+\z}{1-\z}+\frac{ix^2\dot\g}{8\g}
\right)
\ee
where $\z=z\frac{\bar\e}{\e}\,$.

We notice that since the series expansion (\ref{csf}) of the time-dependent CS
(\ref{psiz}) in terms of the time-dependent basis
$\left\{\psi_n(x,t)\right\}$
is exactly the same as the similar expansion for the
time-independent Hamiltonian, all properties of such states known
for the stationary case take place for the non-stationary one.
In particular, they realize the resolution of the identity
operator
and, hence, one can map any square integrable on the positive
semiaxis function to a function holomorphic in the unit disc
getting in such a way a holomorphic representation.
We will not go in  details for this case since they are very
well-known \cite{Per}.

\subsection{Discussion}

\begin{figure}[th]
\label{fig1}
\begin{minipage}{8cm}
\begin{flushright}
\epsfig{file=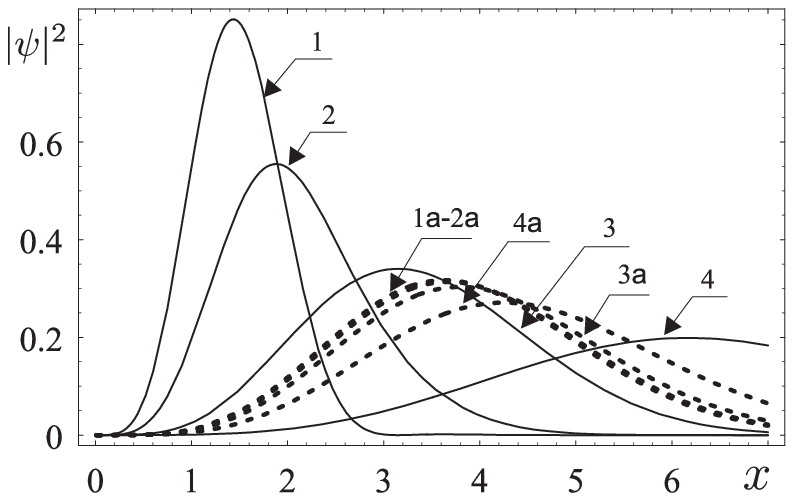, width=6cm}
\caption{\small
Comparison between probability distribution for Barut-Girardelo
and Perelomov CS for time moments $t=0$: $1$
 and $1a$; $t=0.5$: $2$ and $2a$; $t=1$: $3$ and $3a$; $t=2$: $4$
 and $4a$.}
 \end{flushright}
\end{minipage}
\begin{minipage}{8cm}
\hspace{6em}
\epsfig{file=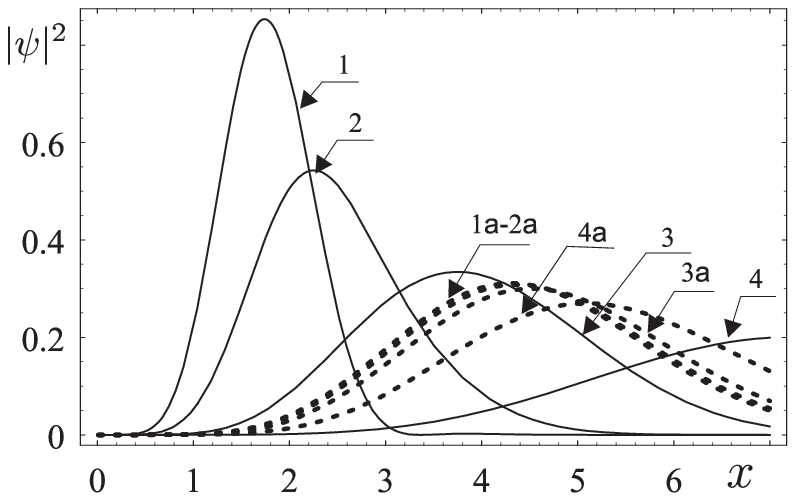, width=6cm}
\caption{\small
Comparison between probability distribution for Darboux transformed
 Barut-Girardelo and Perelomov CS for time moments $t=0$: $1$
 and $1a$; $t=0.5$: $2$ and $2a$; $t=1$: $3$ and $3a$; $t=2$: $4$
 and $4a$.
}
\end{minipage}
\end{figure}

Since the parameters $z$ and $\l$ labelling the CS
fall into different domains, it is not straightforward
 to make a comparison between them.
 One of the possibilities could be to choose these
parameters such that the mean value of an operator is the same for
both states.
As a particular example we take $\omega =0$ (free particle)
and $g=2$ (orbital quantum number $\ell=1$).
In this case $\varepsilon=\frac{1}{\sqrt 2}(t+i)$ and
we have chosen real values for  $z$ and
$\l$ such that the mean value of $k_0$ coincides with its value in
the first excited state (\ref{psin}) which is $k+1$.
In this case $z=(2k+1)^{-1/2}$ and $\l\simeq 1.021$. With these
values of $z$ and $\l$ we plotted the squared modulus of
$\psi_z(x,t)$ (\ref{psiz}) and $\psi_\l(x,t)$ given by
(\ref{psimu}) at $\mu=\sqrt{\l /2}$ in fig. 1.
It is clearly seen from the figure that
at the
initial time moment the
Barut-Girardello  CS are much more localized then Perelomov CS
but the latter are much more stable in time.
 At times
grater then 2 they are already very spread whereas the Perelomov
CS maintain almost the same localization as at the
initial time moment. In the next section we shall show that
the Darboux transformation
keeps the
different time behavior of two types of states practically
unchanged.

\section{Darboux transformation of coherent states}

Time-dependent Darboux transformation \cite{BSrev} can create real
potential differences only if the \Sc equation has at least one
solution $u=u(x,t)$ satisfying the {\it reality condition}
$\left( \log u/\overline u\right) _{xxx}=0$.
Solutions (\ref{psiPQ}) satisfy this condition for any real
function $Q$. So, taking different real solutions of equation
(\ref{Q}) one can, in general, obtain a two-parameter family of
exactly solvable partners for $h_0$ but only the functions
$u=\psi_n(x,t)$
(\ref{psin}) produce potential differences expressed in terms of
elementary functions. Between square integrable solutions only
$\psi_0(x,t)$ is nodeless.
Unfortunately, it produces a
new potential of the same kind as $V_0$ only with a different value
of $g$ (shape invariance at the time-dependent level).
 So, to get essentially new potentials we have to keep in
(\ref{psin}) only upper sign.
For $\a<-3/2$ any nodeless function
(\ref{psin}) gives rise to
a new potential.
For simplicity we will
consider here only the case $\a>0$ when all square integrable
solutions for the transformed Hamiltonian $h_1=-\p_x^2+V_1(x,t)$
can be obtained by acting on corresponding
solutions of the initial equation with the
transformation operator
\be\label{L}
L=L_1(t)[\partial _x-u_x(x,t)/u(x,t)]\qquad
L_1\left( t\right) =
\exp [\, 2\int dt \mathop{\rm Im}\left( \ln u\right)_{xx}\,] \,.
\ee
The potential $V_1(x,t)$ is expressed in terms of the same
function $u$:
\[
 V_1(x,t)=V_0(x,t)+A(x,t)\qquad
 A(x,t)=-[\ln \left| u(x,t)\right| ^2]_{xx}\,.
\]
Taking one of the function (\ref{psin}) with $n=m$
(which is supposed to be fixed from now on)
one gets the potential difference $A(x,t)=A_m(x,t)$:
\bea
A_m(x,t)=
\frac{4k-1}{x^2}+
\frac 18\left( \frac{xL_{m-1}^{2k}\left( z\right) }%
{\gamma L_m^{2k-1}\left( z\right) }\right) ^2
-\frac{x^2L_{m-2}^{2k+1}\left( z\right) +
4\gamma L_{m-1}^{2k}\left( z\right) }{8\gamma ^2L_m^{2k-1}\left( z\right) }
 -\frac 1{4\gamma }
\quad  \\ z=-\frac{x^2}{8\gamma }\,.\nonumber
\eea
The function $L_1(t)$ is determined by (\ref{L}) up to a
constant which we fix to simplify subsequent formulas so that
$L_1=\sqrt{2\g}$ and the explicit expression for the
 transformation operator (\ref{L}) is
\be\label{Lexpl}
L=\sqrt{2\g}\,[\,\p_x-\frac{x}{8\g}-\frac{4k-1}{2x}-
\frac{ix\dot\g}{4\g}-
\frac{xL^{2k}_{m-1}}{4\g L^{2k-1}_{m}}\,]\,.
\ee
Normalized to unity solutions for the Hamiltonian $h_1$ are:
$\vfi_n=N_{1n}L\psi_n$, $N_{1n}=(n+2k+m)^{-{\frac 12}}$.

Operator $L$ and its formally adjoint $L^+$ factorize the symmetry
operator $k_0$: $L^+L=k_0+k+m$. The opposite superposition
$LL^+$ is a symmetry operator for the transformed \Sc equation. If
we denote $p_0=LL^+-k-m$ then the functions $\vfi_n$ are
eigenfunctions of $p_0$ and the spectrum of $p_0$ is identical to
the spectrum of $k_0$. Using $k_\pm$ and transformation operators
$L$ and $L^+$ one can construct other symmetry operators for the
transformed equation: $p_\pm=L\k_\pm L^+$. They are ladder
operators for the functions $\vfi_n$:
\[
p_\pm\vfi_n(x,t)=
(N_{1n}N_{1(n\pm 1)})^{-1}c_n^\pm\vfi_{n\pm 1}(x,t)\,.
\]
We notice now that this Hamiltonian
gives us an example
of the \Sc equation,
symmetry operators of which
do not close a Lie algebra but satisfy a polynomial algebra
\[
[p_0,p_\pm ]=\pm p_\pm
\]
\[
[p_-,p_+]=2[k(1-k)+2p_0(k+m)+2p_0^2](p_0+k+m)\,.
\]
Similar algebra was previously obtained for the time-independent
singular oscillator in \cite{SJMP}.

\subsection{Transformation of Barut-Girardelo coherent states}

To obtain CS for the Hamiltonian $h_1$
we act with $L$ given in  (\ref{Lexpl}) on CS
$\psi_\l$: $\vfi_\l=N_{1\l}L\psi_\l$. The factor $N_{1\l}$ being
calculated from the formula
$N_{1\l}^{-2}=\la\psi_\l|k_0+k+m|\psi_\l\ra$
and (\ref{E0})
guarantees the normalization of the states $\vfi_\l$ to unity.
Their series expansion in terms of the basis $\left\{\vfi_n\right\}$ can be found by
acting with the same operator on the series (\ref{fcs}):
\be
\vfi_\l=N\sum_{n=0}^\infty b_n\l^n\vfi_n\qquad
b_n=a_n(n+2k+m)^{\frac 12}(2k+m)^{-\frac 12}
\ee
where $N=(2k+m)^{\frac 12}N_{1\l}N_{0\l}$.

The states $\vfi_\l=\vfi_\l(x,t)$ thus obtained may be interpreted as
coherent states if they admit the resolution of the identity
operator
\be\label{1r}
\int |\vfi_\l\ra\la\vfi_\l |d\tilde\rho(\l)=1\,.
\ee
Now we proceed to find the measure
$\tilde\rho(\l)$.
We will look for the function $\tilde\rho(\l)$ depending only on
the absolute value of $\l$, $|\l|=\sqrt x$:
$d\tilde\rho =\frac 12h\left( x\right)dxd\phi \,$; the function $h(x)$ is to
be determined.
Therefore, it is
convenient to use polar coordinates in the complex plan of the
variable $\l$,  $\lambda =\sqrt{x}\exp \left( i\phi \right)$.
After being integrated over the variable $\phi$ equation (\ref{1r})
yields
\begin{equation}\label{X}
1=\sum_{n=0}^\infty \frac {\pi(n+2k+m)} {n!\Gamma \left( n+2k\right) }\int_0^\infty
x^{n+\alpha /2-1/4}I_{\alpha -{\frac 12}}^{-1}\left( 2\sqrt{x}\right) h_0\left( x\right)
dx\left| \vfi_n\rangle \langle \vfi_n\right| \,.
\end{equation}
From here it follows that if equation
\begin{equation}\label{Ih}
\frac{\pi(n+2k+m)}{\Gamma \left( n+1\right) \Gamma \left( n+2k\right) }\int_0^\infty
x^{n+\alpha /2-1/4}I_{\alpha -{\frac 12}}^{-1}\left( 2\sqrt{x}\right) h_0\left( x\right)
 dx=1
\end{equation}
is satisfied, than  (\ref{1r}) will also take place because
of the completeness of the system $\left\{\vfi_n\right\}$.
If now we rewrite (\ref{Ih})  as
\begin{equation}\label{Xn1}
\int_0^\infty x^n\Phi \left( x\right) dx=
\frac{\Gamma \left( n+1\right) \Gamma \left(n+2k\right)}%
{\left(n+2k+p\right)}
\end{equation}
where
\begin{equation}\label{Phi}
\Phi \left( x\right) =
\left| N_{0\lambda }N_{1\lambda }\right| ^2h\left( x\right)
\end{equation}
we recognize in it a problem of moments on semiaxis
(see e.g. \cite{Akhieezer}).
To solve this problem we are using the following integral
\cite{BELapl}
\begin{equation}
\int_0^\infty x^nf\left( x\right) dx=
\Gamma \left( n+1\right) \Gamma \left(
n+2k\right)
\label{Xn0}\end{equation}
where $f(x)=2x^{k-{\frac 12}}K_{2k-1}(2\sqrt{x} )$.
It is not difficult to get the expression for
 $\Phi(x)$ in terms of $f(x)$
\begin{equation}\label{Fiy}
\Phi \left( x\right) =
x^{m+2k-1}\int_x^\infty y^{-2k-m}f\left( y\right)dy \,.
\end{equation}

Indeed, first we notice that since $k>\textstyle{\frac 12}$ we have
 $x\Phi \left( x\right) \rightarrow 0$ when $x\rightarrow 0$.
 Therefore, the integration in  (\ref{Xn1}) by parts under the
 condition (\ref{Xn0}) yields just the right hand side of
 (\ref{Xn1}) meaning that equations (\ref{Fiy}) and (\ref{Phi})
 define the measure $\tilde\rho(\l)$.

 The resolution of identity  (\ref{1r}) permits us to construct a
 holomorphic representation for the transformed Hamiltonian $h_1$.
 The Fourier coefficients $\left\{ c_n\right\}$ of a function
  $\varphi(x,t)$
   over the basis $\left\{\varphi_n(x,t)\right\}$
  gives us the same function in CS representation:
  $\varphi ^c\left( \lambda \right) =
  \langle \varphi _{\overline{\lambda }}|\varphi
\rangle =N\varphi \left( \lambda \right)$
and the function
$\varphi \left( \lambda \right) =\sum_{n=0}^\infty b_nc_n\lambda ^n$
is the holomorphic representative of $\vfi_\l(x,t)$.
Now one can define a new inner product
$$
\langle \varphi _1(\l)|\varphi _2(\l)\rangle =
\int \langle \varphi _1|\varphi
_{\overline{\lambda }}\rangle \langle \varphi _{\overline{\lambda }}|\varphi _2\rangle
d\tilde\rho \left( \lambda \right) =
\int \left| N\right| ^2\overline{\varphi }_1\left( \lambda \right)
\varphi _2\left( \lambda \right) d\tilde\rho \left( \lambda \right)
$$
in the space of holomorphic functions,
which gives us a holomorphic representation of states and
operators different from that discussed in section 2.2.

It is not difficult to see that
after the Darboux transformation
any basis function
$\psi_n\left( \lambda \right)$
goes to
$\vfi_n(\l)=(n+2k+m)^{\frac 12}(2k+m)^{-\frac 12}\psi_n(\l)$.
Therefore, if we want for  the functions
$\psi _n\left( \lambda \right)$ and
$ \varphi _n\left( \lambda \right)$
 to be related
by the Darboux transformation,
we have to put
$$\vfi_n(\l)=(n+2k+m)^{-\frac 12}L(\l)\psi_n(\l)\qquad
\psi _n(\l) =(n+2k+m)^{-\frac 12}L^+(\l)\vfi_n(\l)\,.$$
This gives us the holomorphic representation of the Darboux
transformation operators
$$L(\l)=(m+2k)^{-\frac 12}\left[k_0(\l) -k-m\right]\qquad
L^{+}(\l) =(m+2k) ^{\frac 12}\,.
$$

\subsection{Transformation of Perelomov coherent states}

Once again we act with the transformation
operator $L$ given in (\ref{Lexpl}) but now on $\psi_z(x,t)$
(\ref{psiz})
to get the Darboux transformed Perelomov CS,
$\vfi_z(x,t)=N_{1z}L\psi_z(x,t)$. Normalization constant is easily
calculated using the equation
$\la L\psi_z|L\psi_z\ra=\la\psi_z|L^+L\psi_z\ra$ and the
factorization property of the transformation operators,
$N_{1z}^{-2}=m+2k(1-|z|^2)$. Their Fourier series in terms of the
basis $\left\{\vfi_n(x,t)\right\}$ is
\bea\label{fiser}
\vfi_z(x,t)=N_z\sum_{n=0}^\infty b_nz^n\vfi_n(x,t)\\
N_z=N_{0z}N_{1z}(2k+m)^{\frac 12}\,\quad
b_n=a_n(n+2k+m)^\frac 12(2k+m)^{-\frac 12}\,.\nonumber
\eea
Here also the series (\ref{fiser}) is exactly the same
which has previously been obtained for the time-independent
oscillator \cite{SJMP}. We have already found \cite{SJMP}
the measure, which realizes
the resolution of the identity operator in terms of $\vfi_z$,
constructed new Holomorphic representation for the operators and
states, found the K\"ahler potential
and symplectic $2$-form meaning  that we
 obtained a classical
mechanics, which being quantized {\sl \`{a} la} Berezin
gives us back the holomorphic
representation of the quantum system.
This procedure can be considered as the one giving rise to
a classical counterpart of the Darboux transformation
valid both for time-dependent and time-independent cases.

\subsection{Discussion}

\begin{figure}[th]
\label{fig1}
\begin{minipage}{8cm}
\begin{flushright}
\epsfig{file=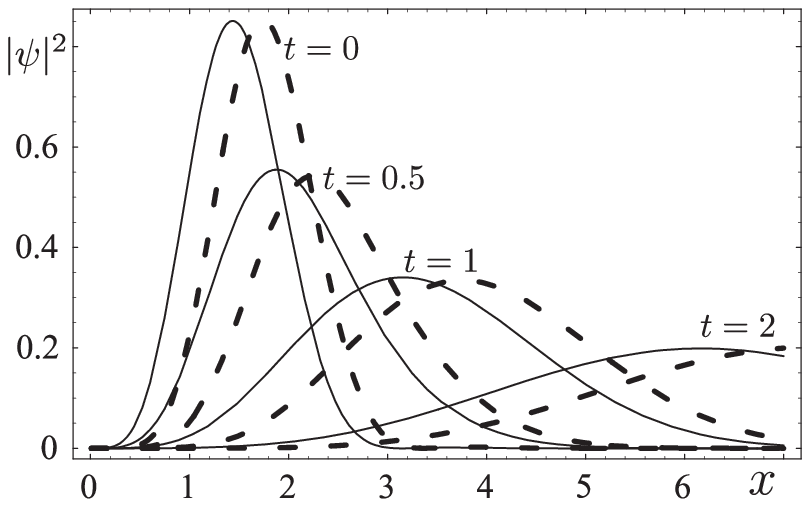, width=6cm}
\caption{\small
Comparison between probability distribution for Barut-Girardelo
 CS before (solid line) and after (dashed line) Darboux transformation. }
 \end{flushright}
\end{minipage}
\begin{minipage}{8cm}
\hspace{6em}
\epsfig{file=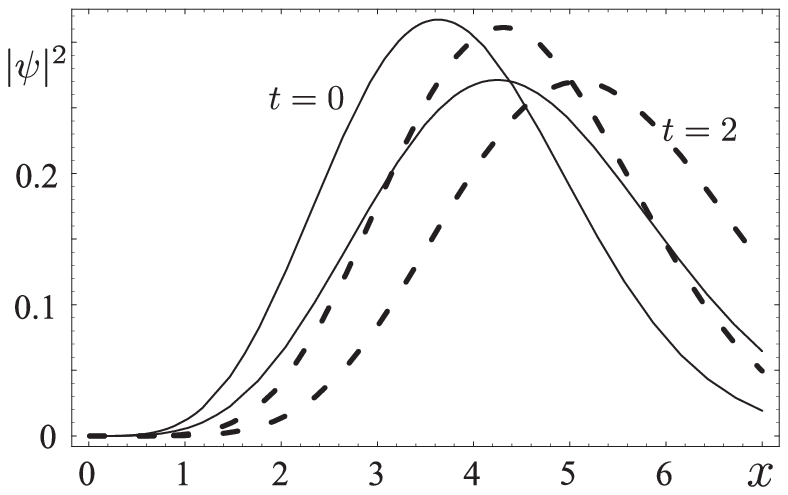, width=6cm}
\caption{\small
Comparison between probability distribution for Perelomov
 CS before (solid line) and after (dashed line) Darboux transformation.
}
\end{minipage}
\end{figure}

We would like to point out that if the \Sc equation has a symmetry
algebra, this property is usually lost after the Darboux
transformation. Nevertheless, if it has ladder operators, the
transformed equation may also have them. In such a case it is
possible to look for eigenstates of the annihilation operator and
call them coherent states. For the case of the time-independent
 harmonic oscillator potential this approach was realized in
 \cite{David}. Since ladder operators have now two derivative orders
 more with respect to the initial ladder operators,
  the differential equation they satisfy is rather complicated
 \cite{Spiridon} which makes difficult studying such states.
 Our approach has an advantage that the transformation operator
 (\ref{Lexpl})
is a
  simple first order differential operator. So, it is very easy
 to operate with it.
 Moreover, in such a way one can get different systems of CS if
 they are available for the initial Hamiltonian.
Usually different systems of CS exhibit  different properties
\cite{opt,AC}.
We conjecture that
Darboux transformation approximately preserves different
behavior of different CS.
To support this conjecture we plotted the transformed
Barut-Girardelo CS together with the transformed Perelomov CS on Fig. 2.
From the first sight the difference between Figs. 1 and 2 is
practically invisible. To show it better we plotted
Barut-Girardelo CS (solid line)
together with their transformed version (dashed line) on Fig.
3. Fig. 4 shows the Perelomov CS before (solid line) and
after (dashed line) the Darboux transformation.
It
is clearly seen from these figures that for both cases the
Darboux transformation results mainly in a displacement of the
curve while its shape is very little affected.

\section{Conclusion}

We have shown that acting with the Darboux transformation operator
on the known CS of the time dependent singular oscillator gives us
the states with the similar properties.
 Thus, the ones obtained
from Barut-Girardelo CS may be called  Barut-Girardelo-like CS
while the others, which are produced using Perelomov CS, may be
called Perelomov-like CS. Each system of CS admits a resolution of
the identity operator which makes it possible to construct
different holomorphic representations. A particular example of a
free particle in $p$-state ($\ell =1$) shows that
Barut-Girardelo-like CS are well localized at the initial time
moment while Perelomov-like CS are more stable with time
evolution. Such a behavior is a reflection of the similar behavior
of corresponding states before the transformation. Therefore we
hope that the new systems of CS may be useful in similar applications
where the known systems have proven to be helpful.

\section*{Acknowledgment}

The work is partially supported by the
President Grant of Russia 1743.2003.2,
the Spanish
Ministerio de Education, Cultura y Deporte Grant SAB2000-0240 and
the Spanish MCYT and European FEDER grant BFM2002-03773.



\section*{References}

\begin{thebibliography}{99}

\bibitem{CS}
Klauder J R and Skagerstam B S 1985 {\sl Coherent states:
applications in physics and mathematical physics} (Singapore: World
Scientific);\\
Malkin I A and Man'ko V I 1979 {\sl Dynamical Symmetries and
Coherent States of Quantum Systems} (Moscow: Nauka);\\
Nieto M and Simmons L 1979 {\sl Phys. Rev.} D {\bf 20} 1321;
Beribe-Lavzier Y and Hussin V 1993 {\sl J. Phys. A: Math. and Gen.}
{\bf 26} 6271


\bibitem{Per}Perelomov A M 1986
{\sl Generalized coherent states and their
applications} (Berlin: Springer)

\bibitem{GK}Gazeau J P  Klauder J R 1999
{\sl J. Phys. A: Math. and Gen.}  {\bf 32} 123

\bibitem{SSP}
Samsonov B F 1998 {\sl JETP} {\bf 114} 1930;\\
Samsonov B F 2000 {\sl J. Phys. A: Math. and Gen.}
{\bf 33} 591

\bibitem{BVM}Brif C, Vourdas A and Mann A 1996
 {\sl J. Phys. A: Math. and Gen.} {\bf 29} 5873

\bibitem{SJMP}Samsonov B F 1998 {\sl J. Math. Phys.} {\bf 39} 967

\bibitem{DCS}
 Samsonov B F 1996 {\sl Phys. Atom. Nucl} {\bf 59} 753

\bibitem{BSDCS}Bagrov V G and Samsonov B F 1996
{\sl J. Phys. A: Math. and Gen.} {\bf 29} 1011;\\
Bagrov V G and Samsonov B F 1996 {\sl JETP} {\bf 109} 1105

\bibitem{BSSO}Bagrov V G and Samsonov B F 1998
{\sl Izv. Vuz. Fiz.} ({\sl Russ. Phys. J.}) {\bf 41(3)} 46

\bibitem{mp}
Hartmann A 1972 {\sl Theor. Chim. Acta} {\bf 24} 201;\\
Chumakov S M, Dodonov V V and Man'ko V I 1986
{\sl J. Phys. A: Math. and Gen.} {\bf 19} 3229;

\bibitem{opt}Dodonov V V, Man'ko V I, Man'ko O V and
 Rosa L 1994 {\sl phys. Lett} A {\bf 185} 231

\bibitem{mph}
Calogero F 1971 {\sl J. Math. Phys.} {\bf 10} 2191;\\
Sutherland B 1971 {\sl J. Math. Phys.} {\bf 12} 246

\bibitem{DMR}Dodonov V V, Manko V I and Rosa L 1998
{\sl Phys. Rev.} A {\bf 57} 2851

\bibitem{M}Miller W (Jr) 1977
{\sl Symmetry and separation of variables}
(Massachusetts: Addison-Wesley)

\bibitem{AC} Agraval G S and Chaturvedi S 1995
 {\sl J. Phys. A: Math. and Gen.} {\bf 28} 5747

\bibitem{BG}Barut A O and Girardelo L 1971
{\sl Commun. Mayh. Phys.} {\bf 21} 41

\bibitem{inv}
Lewis H R and Riesenfeld W B 1969
{\sl J. Math. Phys.} {\bf 10} 1458;\\
Dodonov V V, Manko V I in 1982
{\sl Group theoretical methods in physics} {\bf 1}
Markov M A, Man'ko V I and Shabad A E (Eds)
(Chur: Harward Academic) 591

\bibitem{Beitmen}
Erdelyi A 1953 {\sl Higher transcendental functions}
(New York: McGraw-Hill)

\bibitem{Prudnikov}Prudnikov A P, Brychkov Yu A and Marichev O I
1983 {\sl Integrals and series. Spetial functions} (Moscow: Nauka)

\bibitem{BSrev}Bagrov V G and Samsonov B F 1977
{\sl Phys. Part. Nucl.}  {\bf 28} 374.

\bibitem{David}Fernandez C D J, Hussin V and Nieto L M 1994
 {\sl J. Phys. A: Math. and Gen.} {\bf 27} 3547;\\
 Fernandez C D J, Nieto L M and Rosas-Ortiz O 1995
 {\sl J. Phys. A: Math. and Gen.} {\bf 28} 2693;\\
Rosas-Ortiz O J 1996
{\sl J. Phys. A: Math. and Gen.} {\bf 29} 3281

\bibitem{Spiridon}Spiridonov V 1995
{\sl Phys. Rev. A} {\bf 52} 1909

\bibitem{Akhieezer}Akhieezer N I 1961
{\sl Classical moments problem} (Moscow: Phys.-Math Literature Publishing
Hous)

\bibitem{BELapl}
Erdelyi A 1954 {\sl Tables of integral transformations} V. 1
(New York: McGraw-Hill)

\end{thebibliography}
\end{document}